\newcommand{\be}{\begin{eqnarray}}
\newcommand{\ee}{\end{eqnarray}}
\documentstyle[prc,aps]{revtex}
\begin{document}
\draft
\title{Density and Boundary Effects on Pion Distributions
       in Relativistic Heavy--Ion Collisions}
\author{Alejandro Ayala$^{1}$ and  Augusto Smerzi 
$^{1,2}$\footnote{ Current address: SISSA, Via Beirut 4, 34014, 
Trieste Italy} }

\address{{1)} Department of Physics, University of Illinois at
Urbana-Champaign,
1110 W.Green St. Urbana, IL 61801-3080, USA}

\address{{2)} INFN, Istituto Nazionale di Fisica Nucleare, Italy}
\maketitle
\begin{abstract}

We compute the pion inclusive momentum distribution in a 
heavy--ion collision, assuming thermal equilibrium and accounting for 
boundary effects at the time of decoupling. We calculate the
chemical potential corresponding to an average pion multiplicity in central
collisions and explore the consequences of having the pion system produced 
close to the critical temperature for Bose--Einstein condensation. 

\end{abstract}

\pacs{PACS numbers: 03.65.Pm, 05.30.Jp, 25.75.-q}

In recent years, much experimental effort has been devoted to the
production of matter at high densities and temperatures in relativistic nuclear
collisions. A main goal of such experiments 
is to detect the transition from hadronic matter to the quark gluon 
plasma (QGP). The expectation is that by colliding heavy systems, as opposed
to single hadron--hadron collisions, the chances of producing a highly
compressed state of matter, such as the QGP, are increased. Therefore, any 
experimental signal for the production of QGP is better distinguished 
if we compare similar signatures to those obtained from the collisions of 
smaller systems such as nucleon--nucleon (n--n) collisions.

In this spirit, recent experiments have concentrated their efforts on 
measuring one of those signatures, namely, the inclusive single particle 
momentum distributions for some of the most abundantly produced particles 
after a heavy ion reaction such as pions.

The most striking feature reported~\cite{Strobele}~--~\cite{Barrette2} is an 
enhancement at low as well as at high transverse momentum on the inclusive 
single pion distributions as compared to n--n collisions. This observation has 
sparked a great deal of theoretical effort trying to explain the origin of
such peculiar behavior~\cite{Barz}--~\cite{Gyulassy}. 
Li and Bauer~\cite{Li} have supported the idea that
a superposition of two Maxwell--Bolzmann (M--B) distributions with different 
temperatures might account for the observed behavior. Atwater et 
al~\cite{Atwater} and Lee and Heinz~\cite{Lee} have suggested the possibility 
of transverse flow. Kataja and Ruuskanen~\cite{Kataja}
have fitted a non--zero chemical potential to the Bose--Einstein 
(B--E) distribution. Mostafa and Wong~\cite{Mostafa} stressed the importance 
of properly accounting for boundary effects at freezout. Sollfrank et 
al~\cite{Sollfrank} and G.E. Brown el al~\cite{Brown} (see also 
ref.~\cite{Barrette2}) included the influence of resonant decays into the 
picture.  The effects of Coulomb final state interactions in describing the 
momentum distribution of charged particles has also been pointed 
out~\cite{Gyulassy}.

While the expansion, boundary and Coulomb effects have been amply
discussed in the literature, one of the main features of systems described by
B--E statistics has not been given enough consideration, namely, the fact that
close to some temperature $T_c$, bosons occupy 
predominantly the lowest energy state~\cite{Gerber},~\cite{Pratt}. 
This tendency of bosons to bunch together leads to B--E condensation at $T_c$
and could be responsible for the enhancement of the distribution at low 
$p_t$. It is remarkable, as we will latter show, that the pion multiplicities 
reached in some heavy ion experiments, together with the size of the pion 
source, as measured for example by Hanbury--Brown Twiss (HBT) interferometry, 
yield a value for $T_c$ close to the temperatures which are consistent with
typical abundances~\cite{Braun}.

On the other hand, when the system of pions can be treated as being 
confined just before freezout and the wave functions for the states
satisfy a given condition at the confining boundary~\cite{Mostafa}, the states
form a discrete set. In this case, the density of states contributing to the 
momentum distribution is larger at high $p_t$ compared to a simple B--E 
distribution. This effect, together with an initial transverse flow at 
freezout, could account for the enhancement of the distribution at high $p_t$.

In this paper, we compute the momentum distribution for pions produced
in a relativistic heavy ion collision, assuming that at freezout,
the pions are in thermal equilibrium and incorporating boundary effects. We 
reserve the discussion of possible expansion effects for a following up work.

Recall that a system which consists of a fixed number $N$ of weakly 
interacting, spin zero bosons, in thermal equilibrium at a temperature $T$ is 
described by a grand--canonical ensemble obeying B--E statistics. If $E_i$ 
represents the energy of a single particle state, labeled by $i$,
then the number of particles, chemical potential $\mu$ and 
temperature are related by
\be
   N = \sum_i \frac{1}{e^{(E_i-\mu)/T} -1}\, .
   \label{eq:numpar}
\ee
Since $n(E_i)$ cannot be negative, we have the condition that
$\mu < E_o$, where $E_o$ is the lowest single particle energy state. 
For a fixed number
of particles, $\mu$ is a function of $T$. The peculiar characteristic of 
eq.~(\ref{eq:numpar}) is that when $T$ approaches a certain critical value
$T_c$, $\mu$ approaches the value $E_o$ and thus, this energy state becomes
more populated as $T$ gets closer to $T_c$.

One can estimate the value of $T_c$ by considering the continuum limit of 
eq.~(\ref{eq:numpar}). The result is that for a system of weakly
interacting relativistic bosons, $T_c$ is given implicitly by
\be
   N= \frac{Vm^3}{2\pi^2}\sum_{j=1}^{\infty}
   \left(\frac{T_c}{mj}\right)e^{mj/T_c}K_2\left(\frac{mj}{T_c}\right)
   \label{eq:CT}
\ee
where $V$ is the volume, $m$ is the boson's mass and $K_2$ is the modified 
Bessel function of the second kind and order $2$. However, the above limit
assigns a weight zero to the lowest energy state. Then it is clear that 
when $T\rightarrow T_c$, one needs to return to the discrete picture and
account for the contribution of the individual energy states to quantities 
such as the momentum distribution. For this purpose, we are required to make 
general assumptions about the evolution of the system. 

We assume that the system of pions (of a particular species) produced after
a heavy ion collision is in thermal equilibrium at its time of formation. 
This corresponds to assuming that the total collision rate 
before the time of decoupling is high compared, for instance, to the expansion 
rate.

For a given average multiplicity, we can then use a grand--canonical ensemble 
to describe the statistical properties of the pion system. But statistics alone
is not enough to describe the system's evolution, we also have to account for
the fact that at the time of decoupling, the pion system 
has a finite size and is confined within the boundaries of a given volume.
The shape of this volume is certainly an important issue. Bjorken 
dynamics~\cite{Bjorken}, for instance, implies an overall longitudinal 
expansion and thus that cylindrical geometry is better suited.
Moreover, the time of decoupling is not necessarily the same over the
entire volume~\cite{Bertsch}. However, in order to gain physical insight into 
the problem, we will study the case in which the confining volume has spherical
shape and the decoupling time is unique in the c.m. frame. We thus consider the
following scenario: At the time of decoupling, when strong
interactions have ceased, the system of pions (of a given species) is in thermal
equilibrium and is confined within a sphere of radius $R$ (fireball) as viewed
from the center of mass of the colliding nuclei.

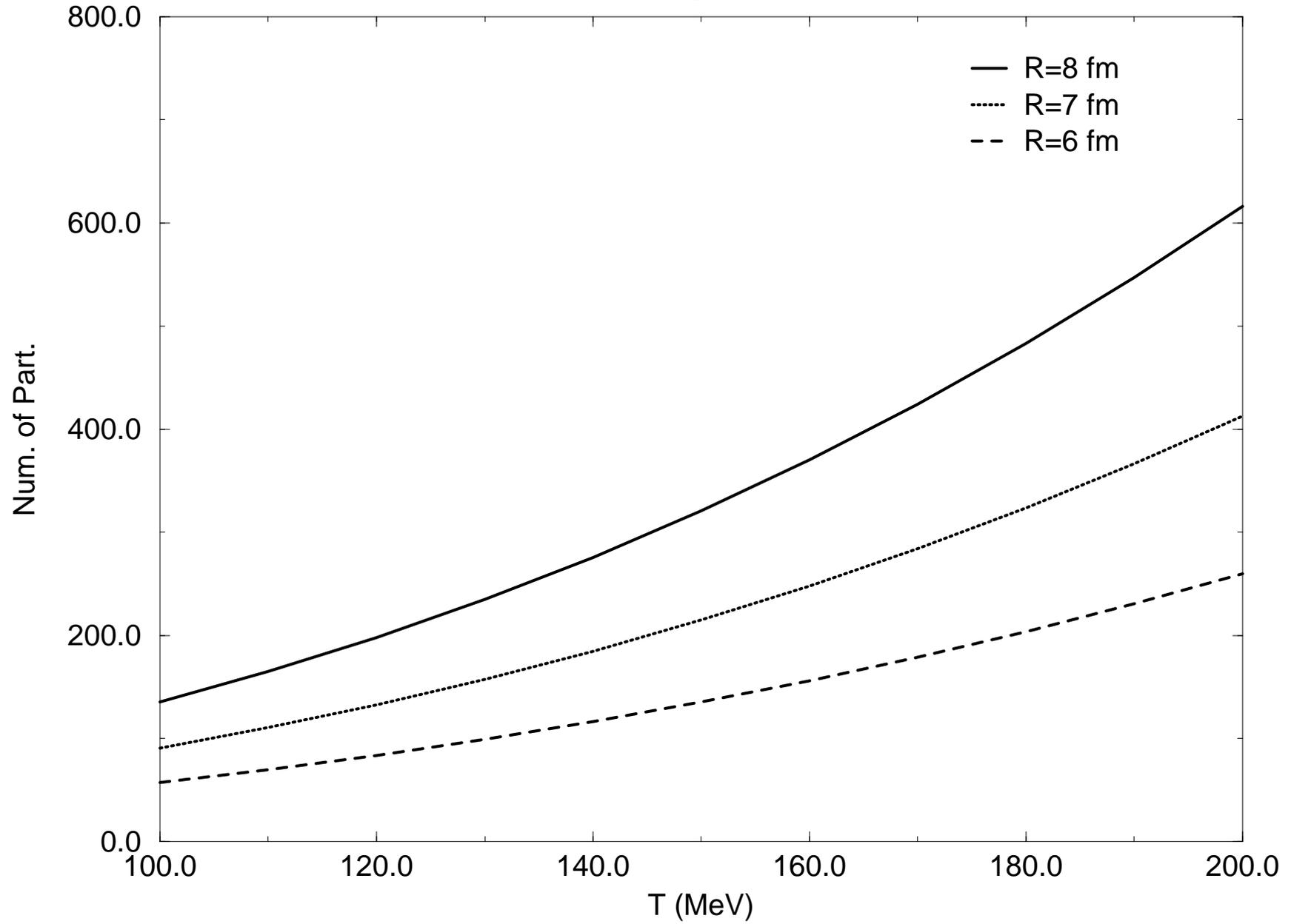
\begin{figure}
\caption{Critical temperature vs. Number of particles for different
values of the radius $R$ at decoupling.}
\label{fig1}
\end{figure}

We start by estimating the critical temperature for the onset of B--E 
condensation. Fig.\ref{fig1} shows plots of $N$ versus $T_c$ computed from 
eq.~(\ref{eq:CT}) for different values of $R$. Recall that the 
total multiplicity in a heavy ion reaction is a function of the invariant 
energy $\sqrt{s}$ in the collision. The multiplicity 
increases logarithmically with $\sqrt{s}$. At AGS energies ($\sqrt{s}\sim$ 
5 A GeV) for example, the average pion multiplicity per event produced in 
central collisions is on the order of 400--500
and thus the number of pions of a particular kind is roughly a third of the 
above. From Fig.\ref{fig1}, we notice that the value of $T_c$ 
for a number of pions of a particular species between 100--200 
is fairly high and decreases when the volume increases. From the discussion 
above, the contribution from the discrete energy states has to be properly 
accounted for. Let us proceed to compute this contribution. 
We first solve for the relativistic wave function corresponding to 
the stationary states of a free particle inside a sphere of radius $R$
\be
   \left( \frac{\partial^2}{\partial t^2} 
   -\nabla^2  + m^2 \right)
   \psi (\vec{r}, t) &=& 0\, .
   \label{eq:expKG}
\ee
We impose the boundary conditions corresponding to a rigid sphere, namely
\be
   \psi(R, t)=0\, ,
   \label{eq:boun}
\ee
to describe the initial particle confinement at freezout. The normalized 
solutions to eq.~(\ref{eq:expKG}), together with the boundary condition, 
eq.~(\ref{eq:boun}), are~\cite{Landau}
\be
   \psi_{klm'}(\vec{r},t)&=&\frac{1}{RJ_{l+3/2}(kR)}
   \left(\frac{1}{rE_{kl}}\right)^{1/2}
   \nonumber \\
   & &Y_{lm'}(\hat{r})J_{l+1/2}(kr)
   e^{-iE_{kl}t}
   \label{eq:solution}
\ee
where $J_n$ is a Bessel function of the first kind and $Y_{lm'}(\hat{r})$ 
are the spherical harmonics. The quantum number $k$ is given by the solution
to
\be
   J_{l+1/2}(kR)=0\, .
   \label{eq:root}
\ee
The energy eigenvalues are related to $k$ by 
\be
   E_{kl}=\sqrt{k^2 + m^2}\, .
   \label{eq:Ekl}
\ee
The normalized contribution to the momentum distribution from the energy 
state with quantum numbers $k,l,m'$ is proportional to the absolute
value squared of the space Fourier transform of eq.~(\ref{eq:solution}) 
\be
   \phi_{klm'}(\vec{p}) = (2E_{kl})
   |\psi_{klm'}(\vec{p})|^2\, ,
   \label{eq:momdensity}
\ee
where
\be
   \psi_{klm'}(\vec{p}) = \frac{1}{(2\pi)^{3/2}}
   \int d^3r e^{-i\vec{p}\cdot\vec{r}}
   \psi_{klm'}(\vec{r})\, . 
   \label{eq:FourTran}
\ee
Due to the azimuthal symmetry of the problem, the wave function in momentum 
space does not depend on the quantum number $m'$ and is a function only of the 
momentum magnitude
\be
   \psi_{klm'}(\vec{p})=
   \psi_{kl}(p)\delta_{m'0}\, .
   \label{eq:del}
\ee
Accordingly, the thermal momentum distribution is given by
\be
   \frac{d^3N}{d^3p} = \sum_{k,l}\frac{\phi_{kl}(p)}
		       {e^{(E_{kl}-\mu)/T}-1}
   \label{eq:dist}
\ee
where $\phi_{kl}(p)$ is given explicitly by
\be
   \phi_{kl}(p)& = &\frac{(2l+1)}{2\pi p}
		    \left[\frac{kJ_{l+1/2}(pR)}{(k^2-p^2)}\right]^2
   \label{eq:expli}
\ee
and $\mu$ is computed from
\be
   N = \sum_{k,l}\frac{(2l+1)}{e^{(E_{kl}-\mu)/T}-1} 
   \label{eq:muexp}
\ee
for fixed N. In general, we can expect an enhancement at high transverse
momentum, relative to a simple B--E distribution, due to the
finite size of the system at freezout, which results in a higher density
of states contributing at high $p$ than at low $p$. The shape of
the distribution at low $p$ should also be affected by a finite chemical 
potential.

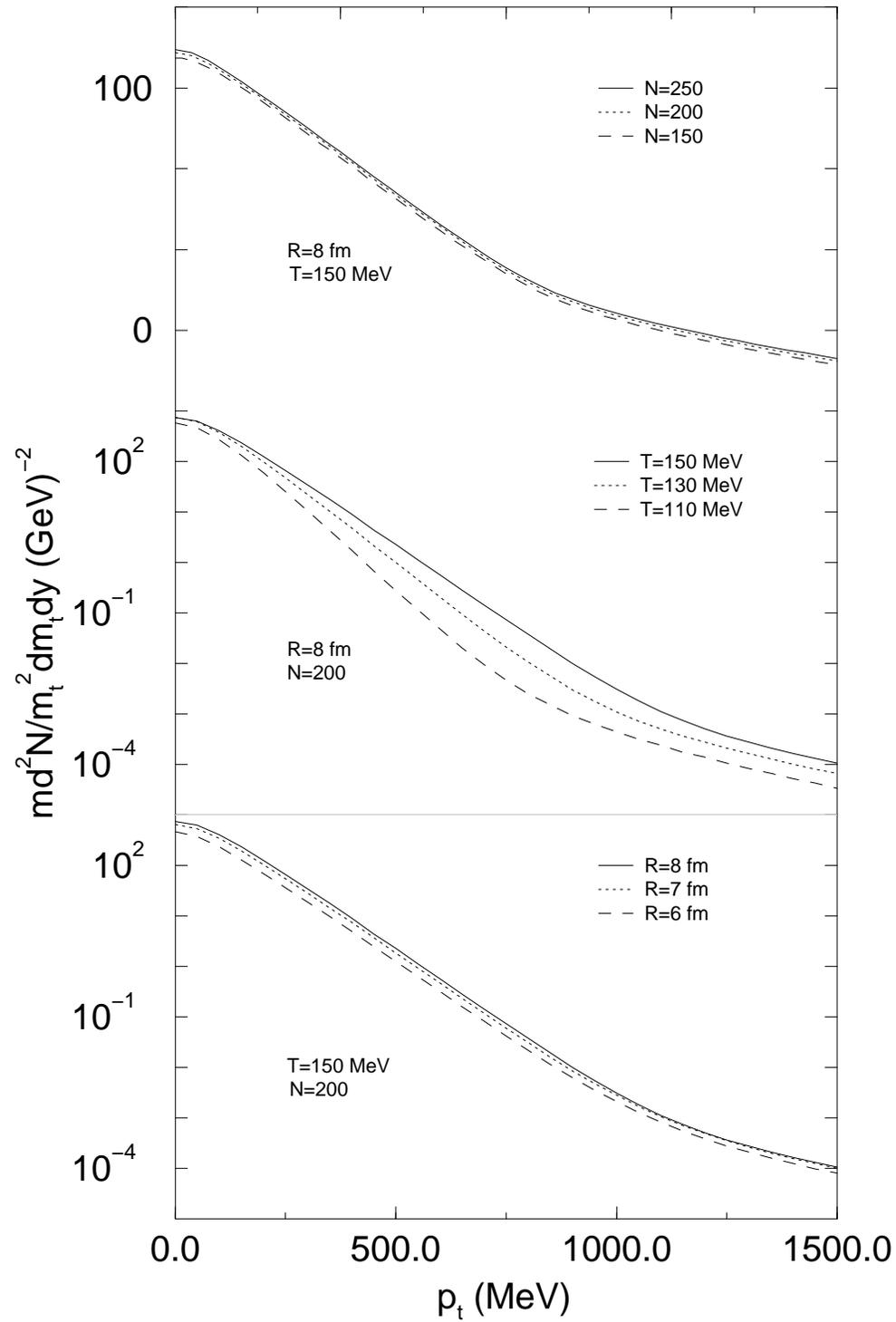
\begin{figure}
\caption{Invariant momentum distribution $md^2N/m_t^2dm_tdy$ at $y=1.4$,
for several values of the parameters $R$, $T$, $N$ and with the computed
values of $\mu$. Notice the bending upwards of the distribution at high $p_t$ 
and the shape at low $p_t$.}
\label{fig2}
\end{figure}

Fig.\ref{fig2} shows the transverse momentum distribution $md^2N/m_t^2dm_tdy$
at fixed $y(=1.4)$, as a function of $p_t$ calculated from eqs.~(\ref{eq:dist}) 
and~(\ref{eq:expli}) for several values of $R$, $T$ and $N$ and
for $\mu$ computed from eq.~(\ref{eq:muexp}). For the chosen set of 
parameters, the main effect is the bending upwards of the distribution at high
$p_t$ relative to a simple exponential with the same temperature. We also
find, in agreement with reference~\cite{Mostafa}, that the distribution
starts deviating from a simple exponential fall off at smaller intermediate 
values of $p_t$ as $R$ decreases.

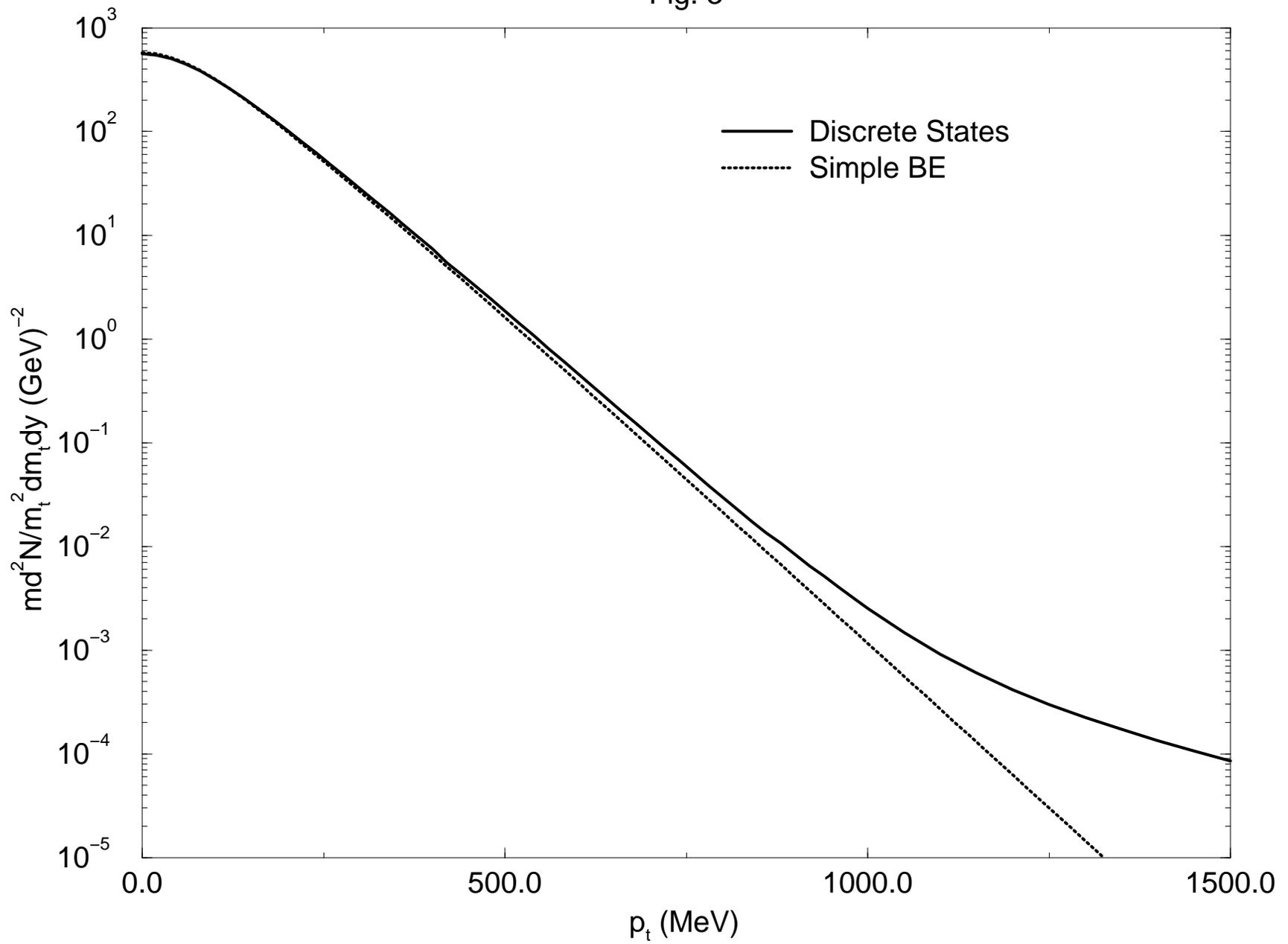
\begin{figure}
\caption{Invariant momentum distribution $md^2N/m_t^2dm_tdy$ at $y=1.4$,
for $R=8$fm, $T=150$MeV and $N=150$ corresponding to a value of $\mu=94.3$MeV.
Shown also is a simple B--E distribution with the same set of parameters 
corresponding to a value of $\mu=67.4$MeV.}
\label{fig3}
\end{figure}

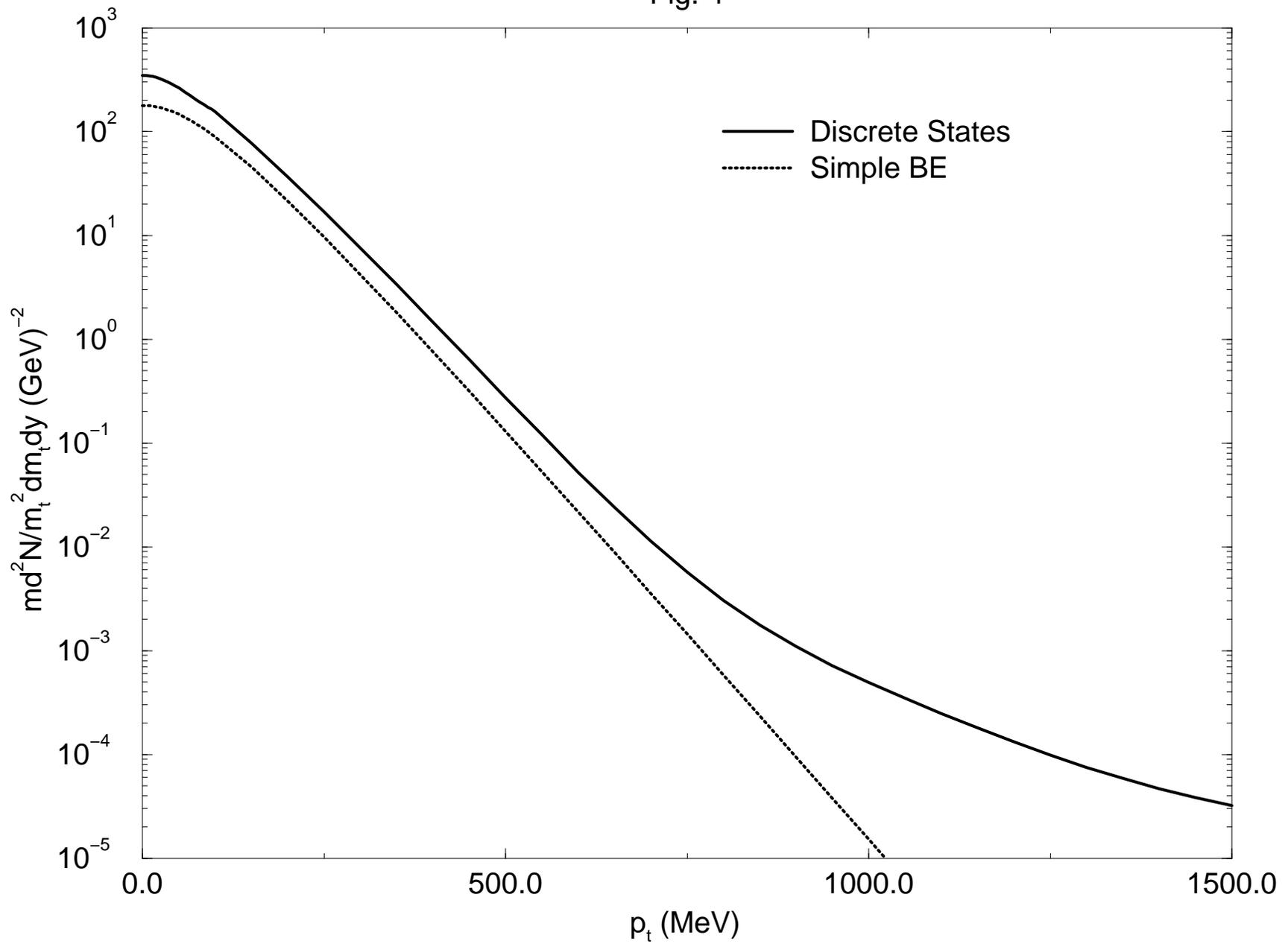
\begin{figure}
\caption{Invariant momentum distribution $md^2N/m_t^2dm_tdy$ at $y=1.4$,
for $R=6$fm, $T=120$MeV and $N=200$ corresponding to a value of $\mu=169.1$MeV.
Shown also is a simple B--E distribution with the same set of parameters
corresponding to a value of $\mu=83.5$MeV.}
\label{fig4}
\end{figure}

Fig.\ref{fig3} shows the distribution for $R=8$ fm, $T=150$ MeV, $N=150$, 
compared to a simple B--E distribution with the same parameters and a chemical
potential corresponding to the same number of particles.
Notice how the distribution in terms of discrete states deviates from the 
simple B--E distribution at high momentum and that, since the parameters are 
far from the critical region for B--E condensation, both distributions coincide
low $p_t$. The situation changes at low $p$ when the parametres
are close to the critical region, this is depicted in fig.\ref{fig4} where
we show the distributions for $R=6$ fm, $T=120$ MeV, $N=200$. The difference 
stems from the way the condensate contribution is included in the discrete 
states picture compared to the continuos case. In the former, the condensate
contribution is spread over different states, although it is mainly 
concentrated in the lowest ones; in the latter, the condensate contribution
is restricted to the lowest energy state, $p=0$. 

In summary, we propose to compute the pion inclusive distribution accounting
for density effects at decoupling by calculating the chemical potential 
corresponding to a given multiplicity. Possible boundary effects can be 
included by a description in terms of a set of discrete states. If the
pion system is produced close to the critical region for B--E condensation,
a description in terms of a discrete set of states lends itself for the
inclusion of the contribution from the lowest energy states. We also
emphasize that the condensation is a high density phenomenon and thus depends
o both, the pion multiplicity and the freezout volume. The presence or 
absence of a condensate could be used to indicate for instance the degree
of transparency in a central collision and have consequences for
correlation experiments.

A.A. thanks G. Baym, J. Kapusta, G. Bertsch and B. Vanderheyden
for insightful discussions and suggestions, the Institute for Nuclear Theory 
at the University of Washington for its hospitality and the Department of 
Energy for partial support during the completion of this work. A.S. thanks 
V.R. Pandharipande for his kind hospitality at the physics department of the 
UIUC. We are indebted to J. Stachel from the E877 collaboration, for 
discussing and making available their data and for very useful comments and 
suggestions. Support for this work has been received in part by the US 
National Science Foundation under grant NSF PHY94--21309.

\centerline{Figure Captions}

\end{document}